\documentclass[preprint,showpacs,preprintnumbers,amsmath,amssymb]{revtex4}

\usepackage{graphicx}% Include figure files
\usepackage{dcolumn}% Align table columns on decimal point
\usepackage{bm}% bold math

\begin{document}

\preprint{to appear in Phys. Rev. Lett.}

\title{Anomalous Threshold Laws in Quantum Sticking}

\author{Dennis P. Clougherty}
%\email{dpc@physics.uvm.edu}
\affiliation{
Department of Physics\\
University of Vermont\\
Burlington, VT 05405-0125}

\date{August 30, 2003}

\begin{abstract}
It has been stated that for a short-ranged surface interaction, the probability of a low-energy particle sticking to a surface always vanishes as $s\sim k$ with $k\to 0$ where $k=\sqrt{E}$.  Deviations from this so-called universal threshold law are derived using a linear model of particle-surface scattering. The Fredholm theory of integral equations is used to find the global conditions necessary for a convergent solution.  The exceptional case of a zero-energy resonance is considered in detail.  Anomalous threshold laws, where $s\sim k^{1+\alpha}, \alpha > 0$ as $k\to 0$, are shown to arise from a soft gap in the weighted density of states of excitations; $\alpha$ is determined by the behavior of the weighted density of states near the binding energy.

\end{abstract}

%\keywords{}%Use showkeys class option if keyword
\maketitle
%\section{introduction}
It has been previously established, both theoretically \cite{berlinsky, kagan, dpc92, carraro92, carraro95, light95, carraro98} and experimentally \cite{yu93}, that at sufficiently low incident energy $E$, the probability of a neutral particle sticking to an insulating substrate vanishes as $s\sim k$ with $k\to 0$ ($k=\sqrt{E}$).  It has been asserted \cite{heller} that this threshold law is independent of the details of the theoretical model and is thus universal.  

It is demonstrated that using a generalization of a linear model of particle-surface scattering previously considered \cite{dpc92}, anomalous threshold laws where $s\sim k^{1+\alpha}, \alpha > 0$ as $k\to 0$ are possible.  These anomalous laws arise with the presence of a soft gap in the weighted density of states of excitations at the binding energy of the particle in the static surface potential.  Thus, the many-body details of the model can indeed alter the threshold law.  

The model is analyzed using the Fredholm theory of integral equations.  There are two advantages of this approach: (1) rigorous conditions on the validity of the solution can be obtained, and (2) the solution is known to be convergent for all coupling strengths.   A general expression for the sticking probability is obtained in terms of the Fredholm determinant.  
Fredholm theory has been previously used in quantum scattering from a static central potential by Jost and Pais \cite{jost51}.  Here, its utility is illustrated for the case of inelastic surface scattering.

The Hamiltonian is taken to be
\begin{eqnarray}
{\cal H}={\cal H}_{p}+{\cal H}_{e}+{\cal H}_{I}
\end{eqnarray}
where
\begin{eqnarray}
{\cal H}_{p}&=&{P^2\over 2m}+V(z),\\
{\cal H}_{e}&=&\sum_i{\omega_i {a_i^\dagger} a_i},\\
{\cal H}_{I}&=&-\gamma\sum_i {W_i\ ({a_i^\dagger} + a_i)\ F(z)} 
\end{eqnarray}
where $F(z)\equiv V'(z)$, and 
${\cal H}_p$ is the Hamiltonian for the particle 
moving in the static potential, $V(z)$. 
${\cal H}_{e}$ is the Hamiltonian for the surface excitations;  
${\cal H}_{I}$ contains the particle--excitation coupling;
$m$ is the particle mass; $\omega_i$ is the frequency of the
excitation in the ith mode; $W_i$ is a generalized mode-dependent coupling; and ${a_i^\dagger}$ and ${a_i}$ are excitation
creation and annihilation operators respectively.  

The Hilbert space is truncated to include only single excitation states.  Thus the statistics of the excitation does not matter here.  The excitation states can be labeled by $|n\rangle$ where $n$ represents the mode that is occupied.  The system wavefunction can be expanded as
\begin{equation}
|\Psi\rangle=\sum_n |n\rangle |\phi_n\rangle
\label{wf}
\end{equation}

The following coupled set of equations result
\begin{eqnarray}
({\cal H}_p-k^2)|\phi_0\rangle&=&\gamma\sum_i M_{0i}|\phi_i\rangle\\
({\cal H}_p-k^2+\omega_i)|\phi_i\rangle&=&\gamma M_{i0}|\phi_0\rangle
\end{eqnarray}
where
\begin{equation}
M_{0i}= {W_i}\ F
\end{equation}
where $F=V'$ and units are chosen so that $E=k^2$.

For the Green's functions to exist for real positive energies, it is neccessary to continue the excitation energies into the lower-half of the complex plane $\omega_j\to \omega_j-i\delta$.  This negative imaginary part of the excitation energies corresponds physically to a reciprocal decay time that is always present in experiment.  It may be thought of as arising from anharmonicity, disorder, or interactions with additional degrees of freedom not explicitly contained in the model.

The coupled set of equations can now be converted into a set of coupled integral equations with the introduction of the outgoing Green's function $G_i={(k^2-{\cal H}_p-\omega_i +i \eta)^{-1}}$.  
\begin{equation}
\phi_i(z;k)=\psi_0(z;k)\delta_{i0}+\gamma\sum_j\int_0^\infty dz'\ K_{ij}(z,z';k)\phi_j(z';k)
\end{equation}
with the kernel $K_{ij}(z,z';k)=-G_i(z,z';k)M_{ij}(z')$.  The incident wave $\psi_0$ is placed in the zero-excitation (or elastic) channel, a boundary condition consistent with a zero-temperature substrate.  Normalization of $\psi_0$ is chosen to give unit flux for the incoming piece.

Such a system of integral equations can be formally solved using Fredholm theory, provided the kernel is sufficiently well-behaved.  The coordinate dependence of $M$ is local, and the kernel can be modified using a transformation for polar kernels \cite{schwinger}:
\begin{equation}
K_{ij}(z,z';k)=-|F(z)|^{\frac{1}{2}}G_i(z,z';k)M_{ij}(z')|F(z')|^{-\frac{1}{2}}
\end{equation}

It is straightforward to show that if the following conditions are satisfied, such a kernel is in the Hilbert-Schmidt class, 
\begin{subequations}
\begin{equation}
\int_0^\infty dz\ z^n |V(z)| <\infty,  \ \ \ n=0,1,2
\end{equation}
\begin{equation}
\int_0^\infty dz\ z^n |F(z)| <\infty,  \ \ \ n=0,1,2
\end{equation}
\begin{equation}
\bigg|\int_0^\infty d\omega\  {\tilde\rho}(\omega){1\over {k^2-\omega+i\delta}}\bigg| < \infty
\label{cond2}
\end{equation}
\end{subequations}
where the {\it weighted} density of states of excitations is given by ${\tilde\rho}(\omega)=|W(\omega)|^2\rho(\omega)$, $\rho(\omega)$ being the density of excitation states. The continuum limit has been taken in Eq.~\ref{cond2}.  These conditions insure that ${\rm Tr}\ K K^\dagger <\infty$. 

The solution of the channel wavefunctions is expressed in terms of the Fredholm resolvent ${\cal R}_{ij}(z,z';k)\equiv {\gamma}{\cal D}_{i j}(z,z';k,\gamma)/\Delta(k,\gamma)$ of the kernel $K$.
\begin{equation}
\phi_i(z)=\psi_0(z)\delta_{i0}+{\gamma\over \Delta(k,\gamma)}\int dz' |F(z)|^{-\frac{1}{2}}{\cal D}_{i 0}(z,z';k,\gamma)|F(z')|^{\frac{1}{2}}\psi_0(z';k)
\end{equation}

Both the numerator of the resolvent ${\cal D}_{ij}$ and the Fredholm determinant $\Delta$ can be expressed as power series expansions in the coupling strength $\gamma$.  It is well-known that these series expansions are absolutely convergent.  Thus the Fredholm method is well suited for investigating the strong coupling regime.

\begin{widetext}
\begin{eqnarray}
{\cal D}_{ij}(z,z')&=&
K_{ij}(z,z')\nonumber\\
&+&
\sum_{n=1}^\infty\sum_{j_n}{(-\gamma)^n\over n!}\int_0^\infty dz_1\dots\int_0^\infty dz_n 
\left|
\begin{array}{llll}
K_{ij}(z,z') & K_{ij_1}(z,z_1) & \cdots & K_{ij_n}(z,z_n) \\
K_{j_1j}(z_1,z') & K_{j_1j_1}(z_1,z_1) & \cdots & K_{j_1j_n}(z,z_n) \\
\vdots & & & \\
K_{j_nj}(z_n,z') & \cdots & & K_{j_nj_n}(z_n,z_n)
\label{dij}
\end{array}\right|\nonumber\\
\\
\Delta&=&
 1\nonumber\\
&+&
\sum_{n=1}^\infty\sum_{j_n}{(-\gamma)^n\over n!}\int_0^\infty dz_1\dots\int_0^\infty dz_n 
\left|
\begin{array}{llll}
K_{j_1j_1}(z_1,z_1) & K_{j_1j_2}(z_1,z_2) & \cdots & K_{j_1j_n}(z_1,z_n) \\
\vdots & & & \\
K_{j_nj_1}(z_n,z_1) & K_{j_nj_2}(z_n,z_2) & \cdots & K_{j_nj_n}(z_n,z_n)
\end{array}\right|\nonumber\\
\label{detseries}
\end{eqnarray}
\end{widetext}

Since $M$ has the form of a border matrix, these series expansions are truncated.  This greatly simplifies Eqs.~\ref{dij} and \ref{detseries} to the following
\begin{eqnarray}
{\cal D}_{i0}(z,z';k,\gamma)&=&K_{i0}(z,z';k)+\gamma\delta_{i0}\sum_m\int dz_1 K_{0m}(z,z_1;k) K_{m0}(z_1,z';k)\\
\Delta(k,\gamma)&=&1-\gamma^2\sum_m \int dz dz'\ K_{0m}(z,z';k)K_{m0}(z',z;k)
\end{eqnarray}

Elements of the S-matrix can be obtained from the asymptotic form of the exact solution.
\begin{subequations}
\begin{equation}
S_{00}={f(k)\over f(-k)}{\Delta(-k,\gamma)\over \Delta(k,\gamma)}
\label{s0}
\end{equation}
\begin{equation}
S_{0n}=-{2i\gamma\sqrt{k k_n}\over\Delta(k,\gamma)}\int dz\ {u(k_n, z)\over f(-k_n)}{u(k, z)\over f(-k)}W_{n} F(z)
\label{sn}
\end{equation}
\end{subequations}
%we use Jost's convention here--not Newton's
where $f(k)$ is the Jost function associated with the potential $V$, $k_n=\sqrt{k^2-\omega_n+i\delta}$, and $u(k,z)$ is the regular solution in the potential $V$ with the condition that $u'(k,0)=1$.

In the absence of sticking, the open-channel S-matrix is unitary.  Conservation of particle flux implies that the deviation from unitarity is the sticking probability $s(k,\gamma)$. Thus
\begin{equation}
{s}(k,\gamma)=1-\sum_{k_n^2\ge 0} |S_{0n}|^2
\label{s}
\end{equation}

The Fredholm determinant is an analytic function of $k$ in the upper-half of the complex $k$-plane.  It is important to note $k=0$ is an exceptional point because of the possibility of a zero-energy resonance in the geometry of surface scattering.  (A centrifugal potential for $\ell\ne 0$, present in scattering from a point source, is absent in the case of surface scattering.) A zero-energy resonance in the potential $V$ permits the Jost function to vanish at zero energy, $f(0)=0$, leading to a simple pole in the Green's function in the elastic channel.  Poles in the inelastic Green's functions $G_m$ have been pushed off the real axis into the lower-half of the complex $k$-plane, with the addition of an imaginary part to the excitation frequencies.  

The Fredholm determinant can be written as
\begin{equation}
\Delta(k,\gamma)=1+\gamma^2 \tau(k)
\label{det}
\end{equation}
where
\begin{equation}
\tau(k)
=-\int dz dz' F(z) F(z') G_0(z,z';k) {\cal I}(z,z';k)
\end{equation}
and
\begin{equation}
{\cal I}(z,z';k)\equiv \int_{0}^\infty d\omega \tilde\rho(\omega) G_0(z,z';\sqrt{k^2-\omega+i\delta})
\end{equation}

Substitution of Eqs.~\ref{s0}, \ref{sn}, and \ref{det} into Eq.~\ref{s} yields
\begin{equation}
s(k,\gamma)={\gamma^2\over |\Delta(k,\gamma)|^2}{\rm Re}\bigg(2 (\tau_c(k)-\tau_c(-k))+\gamma^2(|\tau(k)|^2-|\tau(-k)|^2)\bigg)
\end{equation}
where 
\begin{equation}
\tau_c(k)=-\int dz dz' F(z) F(z') G_0(z,z';k) \int_{k^2}^\infty d\omega \tilde\rho(\omega) G_0(z,z';\sqrt{k^2-\omega+i\delta})
\end{equation}

The low-energy regime is now considered.  
The Green's function $G_0(z,z';k)$ is analytic in the first quadrant of the complex k-plane and can be expanded as a Laurent series about $k=0$.  The series terminates at a simple pole to allow for the possibility of a zero-energy resonance \cite{newton60}.  Thus,  
\begin{equation}
G_0(z,z';k)={g_{-1}(z,z')\over k}+g_{0}(z,z')+g_{1}(z,z') k+\dots
\end{equation}

The discussion now focuses on the case where there is a single bound state in the static potential $V$ at energy $-\kappa_b^2$ with eigenstate $\Phi_b$, real and normalized. 
With
 $\tilde\rho$ suitably smooth,
\begin{equation}
{\rm Im}{\cal I}(z,z';k)\equiv{\cal I}_i(z,z';k)=-\pi\tilde\rho(k^2+\kappa_b^2)\Phi_b(z)\Phi_b(z'), \ \delta\to 0
\end{equation}

Since $g_1$ is pure imaginary,
 for the case where $f(0)\ne0$ ($g_{-1}= 0$), 
 \begin{equation}
{\rm Re} (\tau_c(k)-\tau_c(-k))=2\pi k\tilde\rho(k^2+\kappa_b^2)A_1, \ k\to 0
\end{equation}
where 
\begin{equation}
A_1=\int dz dz' F(z) F(z')\Phi_b(z)\Phi_b(z'){\rm Im}\ g_1(z,z')
\label{a}
\end{equation}

Similarly,
\begin{equation}
\tau(k)\tau^*(k)-\tau(-k)\tau^*(-k)= 4\pi k\tilde\rho(k^2+\kappa_b^2)B_1, \ k\to 0
\end{equation}
where
\begin{eqnarray}
B_1&=&\int dz_1 dz'_1 dz_2 dz'_2 F(z_1) F(z'_1)F(z_2)F(z'_2)
g_0(z_1,z'_1) {\rm Im}\ g_1(z_2,z'_2)\nonumber\\
&\times&(\Phi_b(z_2)\Phi_b(z'_2){\cal I}_r(z_1,z'_1)-\Phi_b(z_1)\Phi_b(z'_1){\cal I}_r(z_2,z'_2))
\label{b}
\end{eqnarray}
and ${\cal I}_r(z,z')={\rm P} \int_{0}^\infty d\omega \tilde\rho(\omega) G_0(z,z';i\sqrt{\omega})$.

Thus, the low-energy sticking probability is of the form
\begin{equation}
s(k,\gamma)={4\pi\gamma^2(A_1+\gamma^2 B_1)\over 1+\gamma^2 C_0+\gamma^4 D_0}k\tilde\rho(k^2+\kappa_b^2) , \ \ \ \ k\to 0
\label{stick3}
\end{equation}
where 
\begin{eqnarray}
C_0&=& -2\int dz dz' F(z) F(z') g_0(z,z'){\cal I}_r(z,z')\\
D_0&=&\int  dz_1 dz'_1 dz_2 dz'_2 F(z_1) F(z'_1)F(z_2)F(z'_2)g_0(z_1,z'_1)g_0(z_2,z'_2)\nonumber\\
&\times&({\cal I}_r(z_1,z'_1){\cal I}_r(z_2,z'_2)+\pi^2\tilde\rho^2(\kappa_b^2)\Phi_b(z_1)\Phi_b(z'_1)\Phi_b(z_2)\Phi_b(z'_2))
\label{d}
\end{eqnarray}

Under typical circumstances, $\tilde\rho(k^2+\kappa_b^2)$ approaches a finite and non-zero limit as $k\to 0$.  Then the ``universal'' sticking law is recovered from Eq.~\ref{stick3} for the case of a short-ranged potential, with $s\sim k$ as $k\to 0$.

Anomalous threshold laws, where $s\sim k^{1+\alpha}$ ($\alpha > 0$), are theoretically possible for cases where the weighted density of states has a soft gap approaching zero at the binding energy.  It is noted that a soft gap in the density of states frequently occurs in the vicinity of a phase transition.  Scattering from a substrate undergoing a at low temperature structural phase transition  might reveal such anomalous laws.  The temperature of the substrate could be used to tune the softening phonon mode.  In the vicinity of a minimum in the dispersion, a soft gap in the density of states appears, with $\rho(\omega)\propto \sqrt{\omega-\omega_0}$ in the neighborhood above the softened frequency $\omega_0$.  For the case of a smooth weighting function $W$, non-vanishing at $\kappa_b^2$,  $\alpha=1$ and $s\sim k^{2}$ as $k\to 0$.

The phonon and roton contributions to sticking have been independently measured \cite{wyatt95, wyatt03} for helium scattering from superfluid $^4$He.  The roton density of states of superfluid $^4$He vanishes as the energy approaches the minimum energy to create a roton.  It is known \cite{carraro92, hijmans} that the static potential can be substantially modified by preparing a film of superfluid $^4$He on a substrate with a van der Waals coefficient different from that of the $^4$He film.  If the depth of the binding energy is adjusted to the roton minimum energy by using the appropriate substrate and film thickness, then it might be possible to measure an anomalous threshold law for sticking by roton creation.  

 Another possibility would be to use a substrate where the optical phonon energy at the zone edge is equal to the binding energy.  While surface preparation could be used to adjust the binding energy, the van Hove singularity of a bulk optical phonon at the zone edge is insensitive to surface perturbations and can provide the necessary soft gap.  Flatt{\' e} and Kohn \cite{flatte} have discussed the general conditions for dominance of inelastic scattering by bulk substrate parameters.

Consider the case of $\tilde\rho(\omega)\sim (\omega-\kappa_b^2)^\eta$ for $\omega\ge \kappa_b^2$ ($\eta >0$).  The sticking probability then behaves as
\begin{equation}
s(k,\gamma)={4\pi\gamma^2(A_1+\gamma^2 B_1)\over 1+\gamma^2 C_0+\gamma^4 D_0}k^{1+2\eta} , \ \ \ \ k\to 0
\label{thresh1}
\end{equation}

The suppression of the weighted density of states at the binding energy results in a  decline in the sticking probability at low energies that is more rapid than would be predicted from the ``universal'' threshold law.  Under the conditions considered, the frequency dependence of the weighted density of states near the binding energy appears in the energy dependence of the sticking probability near zero energy.

While a zero-energy resonance is not a true bound state, it may be present in addition to a bound state.  The presence of the zero-energy resonance affects the energy dependence of the elastic Green's function and alters the form of the sticking probability with coupling strength $\gamma$.

With $f(0)=0$ ($g_{-1}\ne 0$) and $\gamma\ne 0$, at suitably low energies,
\begin{equation}
s(k,\gamma)={4\pi(A_{-1}+\gamma^2 B_{-1})\over \gamma^2 D_{-1}}k^{1+2\eta} , \ \ \ \ k\to 0
\label{thresh2}
\end{equation}
where $A_{-1}$ and $B_{-1}$ are obtained from Eqs.~\ref{a} and \ref{b} by replacing $g_1\to g_{-1}$ and $D_{-1}$ by replacing $g_0\to {\rm Im}\ g_{-1}$ in Eq.~\ref{d}.
The same k-dependence of the threshold laws results as in Eq.~\ref{thresh1}; however, the dependence on coupling strength differs.

Anomalous threshold laws in low energy particle-surface scattering have been shown to result from a soft gap in the weighted density of states at the frequency equal to binding energy.  This provides a counterexample to the claim \cite{heller}  that the ``universal'' threshold law is independent of model details.  Fredholm theory has been used to find an exact solution together with global conditions on the potential and inelastic coupling functions for the validity of the solution.   The solution of the model is valid for arbitrarily large coupling strengths $\gamma$.
A general expression for the sticking probability in terms of the Fredholm determinant has been derived.  The exceptional case of a zero-energy resonance is seen to affect the coupling constant dependence, but not affect the energy dependence of the sticking probability.  

\bibliography{anomaly}
\end{document}